\documentclass[a4paper,11pt]{article}
\pdfoutput=1 
\usepackage{jinstpub} 

\usepackage{subcaption}
\usepackage{xcolor}
\usepackage{physics}
\usepackage{dsfont}
\usepackage{bm}
\graphicspath{
    {./img/}
}

\DeclareMathOperator\sign{sign}

\renewcommand\vec[1]{\boldsymbol{\mathbf{#1}}}

\title{Linear approximation to the statistical significance autocovariance matrix in the asymptotic regime.}

\author{V. Ananiev,}
\author{A. L. Read}
\affiliation{Department of Physics, University of Oslo, Boks 1072 Blindern, Oslo, NO-0316, Norway}

\emailAdd{a.l.read@fys.uio.no}

\abstract{%
Approximating significance scans of searches for new particles in high-energy physics experiments as Gaussian fields is a well-established way to estimate the trials factors required to quantify global significances. We propose a novel, highly efficient method to estimate the covariance matrix of such a Gaussian field. The method is based on the linear approximation of statistical fluctuations of the signal amplitude. For one-dimensional searches the upper bound on the trials factor can then be calculated directly from the covariance matrix. For higher dimensions, the Gaussian process described by this covariance matrix may be sampled to calculate the trials factor directly. This method also serves as the theoretical basis for a recent study of the trials factor with an empirically constructed set of Asmiov-like background datasets. We illustrate the method with studies of a $H \rightarrow \gamma \gamma$ inspired model that was used in the empirical paper.
}

\keywords{Analysis and statistical methods, Simulation methods and programs, Data processing methods}

\begin{document}

\maketitle
\flushbottom

\section{Introduction}

In high energy physics searches for new particles that appear in the data as resonances~\cite{CMSHyy,ATLASHyy},
one usually scans a mass region and hopes to find a peak of high significance at some mass.
The significance at each mass of the scan is generally found by applying Wilks' theorem~\cite{wilks1938}
to the likelihood-ratio test statistic (LRT)~\cite{cowan2011}
for each point,
and results in a field of significances measured across the search region.

While the resonance may appear anywhere in the search region, the analysis
usually targets the highest (local) significance,
which leads to the recurring challenge of estimating the global significance of this
observation.
The necessity of calculating the probability for a background fluctuation to give such a peak of significance anywhere in the search region,
and not simply where the significance is maximal,
is commonly referred to as the look-elsewhere effect (LEE).

There have been a number of studies investigating the LEE, and in our work we
pay particular attention to those describing the significance field with a Gaussian process.
While some studies~\cite{gross2010,vitells2011,davies77}
set the upper bound on the trials factor,
which converts a local p-value into a global one,
and only use a Gaussian process implicitly to link
the low and high significance regions,
other studies~\cite{ananiev2022} require explicit values for
the Gaussian process parameters.

In this paper we establish a chain of lightweight steps
from a non-linear parametric statistical model to the trials factor by estimating the covariance matrix of the significance field.
To construct the estimate involving only one background only fit to the data,
we apply linear expansion to the non-linear background shape.
The way to calculate the covariance matrix starting from a linear model
was briefly discussed by Demortier~\cite[p.~23-33]{CERN-2008-001}.
As part of our work, we give a strict mathematical formulation of the method and demonstrate a practical application of it to non-linear background shapes,
with the estimated covariance matrix serving as a proxy for the straightforward trials factor estimate.

A common input for the methods that quantify the LEE is a set of maximum likelihood fits to
some number of Monte Carlo generated data realizations. They may be used to
estimate the trials factor in the lower significance region, or the covariance
matrix of the Gaussian process itself (the significance autocovariance).
The challenge, then, is to fit enough
datasets to estimate the trials factor with a satisfactory precision, while
keeping the number of fits as small as possible.

In high-energy physics searches for a new particle or a resonance, typically,
the likelihood-ratio test statistic is used to construct the p-value for
each point on a search grid.
In the asymptotic regime, the test statistic follows a $\chi^2$ distribution.

For analyses that use a Gaussian process to model the significance,
the number of degrees of freedom of the test statistic distribution is,
typically, 1. For this case, in Chapter~\ref{sec:method},
we suggest a method to estimate the significance covariance matrix
that makes use of a single background-only fit to the data.

We replace the set of fits that were required in our previous work,
with derivatives of the best-fit-to-the-data background model.
Fortunately, the derivatives can often be extracted from the fit software.

\paragraph{Core assumptions.}\label{par:core-assumptions} In section~\ref{sec:asimov-set-just} we show that three
quite generic requirements:
\begin{enumerate}
    \item the background model should be well approximated by its linear expansion around the best fit parameters,
    \item the assumption that the data can be binned and fluctuations in different bins of the dataset
        are independent,
    \item the fluctuations in each bin follow a Gaussian distribution,
\end{enumerate}
together, are consistent with the assumptions made in the empirical study by Ananiev \& Read~\cite{ananiev2022},
which relied on the additivity (superposition) principle
for the fluctuations to empirically estimate the covariance matrix
of the significances.
We argue, therefore, that this work serves
as a theoretical basis for the method of the set of Asimov background samples introduced in the study,
and at the same time may rely on its validations.

\subsection{Statistical model}

The basic structure of a statistical model commonly used in high-energy physics
experiments that search for a new particle or a resonance was described in
detail in the empirical study~\cite{ananiev2022}.
For the present study, we chose the $H\rightarrow\gamma\gamma$ inspired model as a benchmark,
because it satisfies without approximation the second and third requirements above.

The search is conducted with the likelihood ratio test statistic evaluated
for each point $M$ of the search grid $\mathcal{M}$.

In this binned model,
the expected background $b_i(\vec{\theta})$ has an exponential shape and is used as the null-hypothesis $H_0$.
The shape of the expected signal $\mu s_i(\vec{\theta})$ is Gaussian and together with background $b_i$
forms the alternative $H_1$, expected signal + background estimate:
\begin{align}
    n_i(\mu, \vec{\theta}, M) = \mu s_i(\vec{\theta}, M) + b_i(\vec{\theta}),
\end{align}
where $i$ enumerates bins, $\vec{\theta}$ denotes the vector
of nuisance parameters and $\mu$ is the signal strength parameter.

Generally, in the asymptotic regime (e.g.\ large sample), and neglecting constant terms, log-likelihoods for $H_0$ and $H_1$ may be approximated as follows\footnotemark{}:
\footnotetext{%
    The $H\gamma\gamma$ inspired model assumes Gaussian errors in its definition~\cite{ananiev2022}. The expressions for log-likelihoods (eq.~\ref{eq:loglikelihoods}) in case of this model are, therefore, exact.
}
\begin{align} \label{eq:loglikelihoods}
    &-2\ln \mathcal{L}_0(\mu=0, \vec{\theta}) = \sum_i {\left( \frac{d_i - b_i(\vec{\theta})}{\sigma_i} \right)}^2, \\
    &-2\ln \mathcal{L}_1(\mu, \vec{\theta}, M) = \sum_i {\left( \frac{d_i - b_i(\vec{\theta}) - \mu s_i(M, \vec{\theta})}{\sigma_i} \right)}^2, \nonumber
\end{align}
where $i$ enumerates bins, $M \in \mathcal{M}$ denotes the point in the search region $\mathcal{M}$ of parameters which are not present under the background-only hypothesis, $\vec{\theta}$ are the nuisance parameters, and $d_i$ corresponds to the binned data with errors $\sigma_i$\footnotemark{}.

\footnotetext{%
We have assumed that the errors $\sigma_i$ are independent of the nuisance
parameters $\vec{\theta}$.
With a linear correction to $\sigma_i$ it is still possible to get a closed form expression
for the test statistic and significance. The calculation of the covariance would require sampling toys to average out the fluctuations.
No additional fits would be required, however,
so this may be a potential option for more sophisticated analyses.%
}

\begin{figure}[t]
    \centering
    \includegraphics[width=0.6\linewidth]{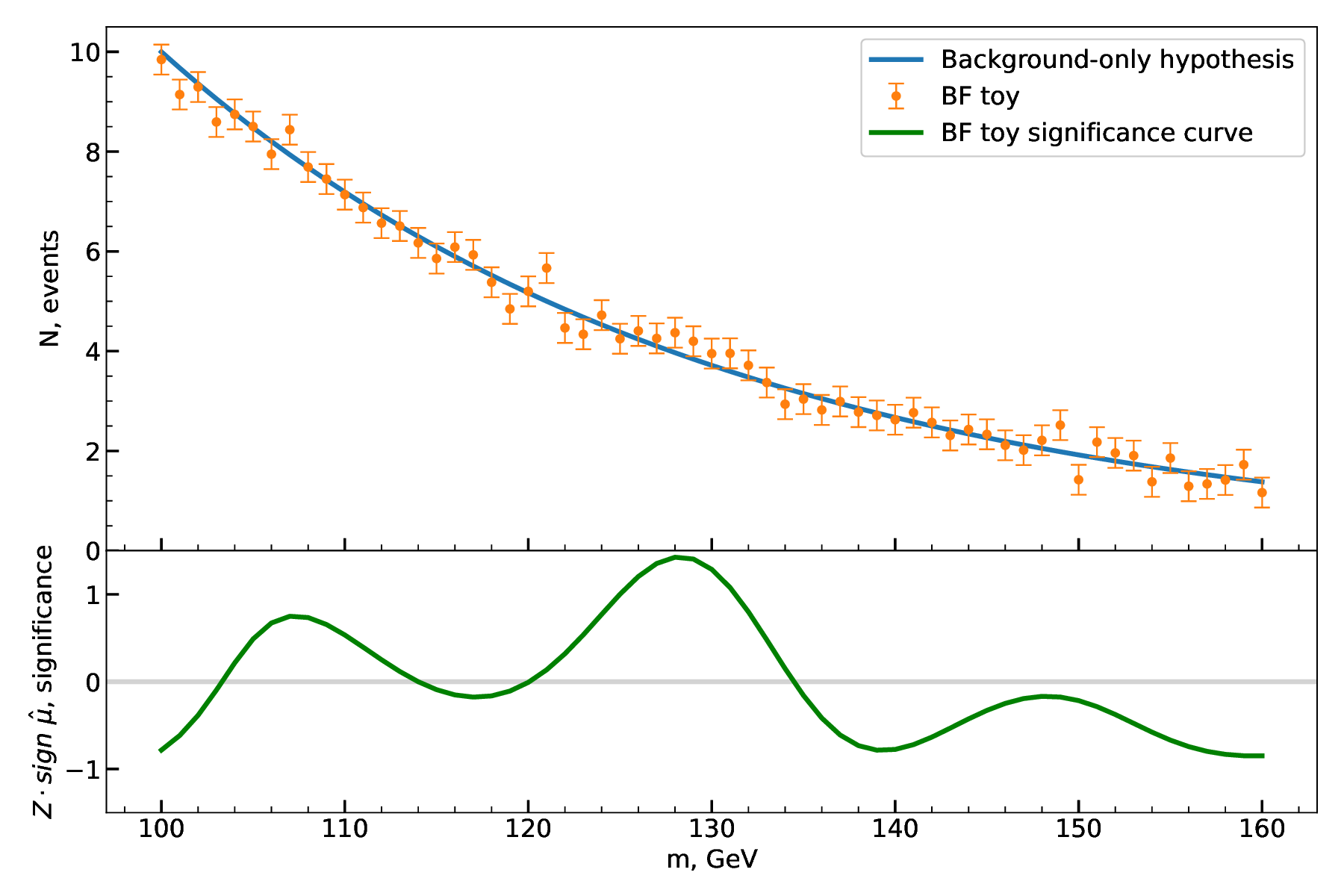}
    \caption{One brute force toy (orange) sampled from a background-only hypothesis (blue) of the $H \rightarrow \gamma\gamma$ inspired model. The signal significance curve is plotted below (green), where each point of the curve corresponds to a different choice of the signal hypothesis.}\label{fig:hyy-one-toy}
\end{figure}

Our goal is to estimate the covariance matrix $\Sigma_{MN}$ of the statistical significances $Z_{M}$ and $Z_{N}$ evaluated at two different points of the search region $\mathcal{M}$:
\begin{align}
    &\Sigma_{MN} = \langle Z_M Z_N \rangle_{d}, \quad M, N \in \mathcal{M}, \\
    &Z_M = \sign(\hat{\mu}) \sqrt{t_{\mu}(M)} \sim \mathcal{N}[0, 1], \label{eq:def-sig} \\
    &t_{\mu}(M) = -2 \ln \frac{\mathcal{L}_0(\mu=0, \vec{\theta}_0)}{\mathcal{L}_1(\hat{\mu}, \vec{\theta}_0 + \vec{\theta}_1, M)} \sim \chi^2_{d.o.f=1},
\end{align}
where $t_{\mu}(M)$ is the likelihood-ratio test statistic (LRT), $Z_M$ is the so-called signed-root LRT,
$\vec{\theta}_0$ are the nuisance parameters that maximize
the background-only likelihood $\mathcal{L}_0$,
$\vec{\theta}_0 + \vec{\theta}_1$ together with the signal strength $\hat{\mu}$
maximize the signal+background likelihood $\mathcal{L}_1$, and $\mathcal{N}[0, 1]$ denotes the standard normal distribution.

To give a feeling of the $H\gamma\gamma$ inspired model, in figure~\ref{fig:hyy-one-toy} we plotted the shape of the background-only hypothesis $b_i$, one sample of data sampled from it, and a corresponding significance curve (eq.~\ref{eq:def-sig}). Notice how clearly visible bumps in the data are reflected in peaks of the significance curve.

We would like to remark that for the signal+background model we are fitting $\vec{\theta}$ as a deviation from $\vec{\theta}_0$.
This is essential for the proper separation of variables in the subsequent calculations.

We assume that the best fit of the background model $b_i$ to the data $d_i$ is available for the study as $b_i(\vec{\hat{\theta}}) = \hat{b}_i$. In order to simplify the notation, we make use of the freedom to choose the reference point for the model parameters $\vec{\theta}$ and define the best fit parameters to be $\vec{\hat{\theta}} = \vec{0}$.

\section{Method}\label{sec:method}

To simplify the notation, we redefine $d_i$, $s_i$ and $b_i$ to include $\sigma_i$:
\begin{align}
    \frac{d_i}{\sigma_i} \mapsto d_i, \quad
    \frac{s_i}{\sigma_i} \mapsto s_i, \quad
    \frac{b_i}{\sigma_i} \mapsto b_i. \label{eq:norm}
\end{align}
The log-likelihoods then become:
\begin{align}
    &-2\ln \mathcal{L}_0 = \sum_i {\left( d_i - b_i(\vec{\theta}) \right)}^2, \\
    &-2\ln \mathcal{L}_1 = \sum_i {\left( d_i - b_i(\vec{\theta}) - \mu s_i(\vec{\theta}) \right)}^2. \nonumber
\end{align}

For every realization of the data, we expect the deviations of the fit parameters $\mu$ and $\vec{\theta}$ from $0$ to be small (in the absence of a signal), and therefore the first-order expansion of $b_i(\vec{\theta})$ and $s_i(\vec{\theta})$ around $\vec{0}$ to be accurate enough.

The log-likelihoods then are:
\begin{align}\label{eq:b_loglike_expansion}
    &-2\ln \mathcal{L}_0 = \sum_i {\left( d_i - \hat{b}_i - \Delta_{i \beta} \theta^\beta \right)}^2, \\
    &-2\ln \mathcal{L}_1 = \sum_i {\left( d_i - \hat{b}_i - \Delta_{i \beta} \theta^\beta - \mu s_i(\vec{0}) \right)}^2, \nonumber
\end{align}
where $\Delta_{i \alpha} = \frac{\partial b_i(\vec{\theta})}{\partial \theta^\alpha} \vert_{\vec{\theta} = \vec{0}}$ is the Jacobian of the best-fit background model and the Einstein summation rule applies to the indices $\beta$.
Since the signal model $s_i$ contributes to the log-likelihoods eq.~(\ref{eq:b_loglike_expansion}) only at lowest order, thus is constant, we simplify $s_i(\vec{0})$ to $s_i$ from now on.

The equations that define optimal values of $\vec{\theta}_0$, $\vec{\theta}_1$, and $\mu$ then are:
\begin{align}
    &\frac{\partial \mathcal{L}_0}{\partial \theta_\alpha} \vert_{\vec{\theta}_0} \propto \\
    &\quad \sum_i (d_i - \hat{b}_i - \Delta_{i \beta} {\theta_0}^{\beta})\cdot \Delta_{i\alpha} = 0, \nonumber \\
    &\frac{\partial \mathcal{L}_1}{\partial \theta_\alpha} \vert_{\vec{\theta}_1, \hat{\mu}} \propto \\
    &\quad \sum_i (d_i - \hat{b}_i - \Delta_{i \beta} ({\theta_0}^{\beta} + {\theta_1}^{\beta}) - \hat{\mu} s_i)\cdot \Delta_{i\alpha} = 0, \nonumber \\
    &\frac{\partial \mathcal{L}_1}{\partial \mu} \vert_{\vec{\theta}_1, \hat{\mu}} \propto \\
    &\quad \sum_i (d_i - \hat{b}_i - \Delta_{i \beta} ({\theta_0}^{\beta} + {\theta_1}^{\beta}) - \hat{\mu} s_i)\cdot s_i = 0. \nonumber
\end{align}

To reduce the number of indices, we rewrite the expressions above with bra-ket notation:
\begin{align}
    &\bra{d -\hat{b}} \Delta =  \bra{\theta_0} \Delta^{\intercal} \Delta, \label{eq:master-1} \\
    &\vec{0} =  \bra{\theta_1} \Delta^{\intercal} \Delta + \hat{\mu} \bra{s} \Delta, \label{eq:master-2} \\
    &\braket{d - \hat{b}}{s} = \bra{\theta_0 + \theta_1} \Delta^{\intercal} \ket{s} + \hat{\mu} \braket{s}{s}, \label{eq:master-3}
\end{align}
where in eq.~(\ref{eq:master-2}) we used eq.~(\ref{eq:master-1}) to cancel the $\vec{\theta}_0$ contribution. We can solve eq.~(\ref{eq:master-1}) and eq.~(\ref{eq:master-2}) for $\vec{\theta}_0$ and $\vec{\theta}_1$ correspondingly:
\begin{align}
   &\bra{\theta_0} =  \bra{d-\hat{b}} \Delta {(\Delta^{\intercal} \Delta)}^{-1}, \label{eq:nu-0} \\
   &\bra{\theta_1} = - \hat{\mu} \bra{s}  \Delta {(\Delta^{\intercal} \Delta)}^{-1}. \label{eq:nu-1}
\end{align}

It is important to mention that, although  $\Delta$ itself is generally singular, the product $\Delta^{\intercal} \Delta$ appears to be a Hessian of $-2\ln \mathcal{L}_1$ with respect to $\vec{\theta}_1$. For the background model best-fit point $\vec{\theta} = \vec{0}$ to be a minimum, it is required that the Hessian be positive definite, thus $\Delta^{\intercal} \Delta$ is invertible.

We substitute eq.~(\ref{eq:nu-0}) and eq.~(\ref{eq:nu-1}) into eq.~(\ref{eq:master-3}) and solve for $\hat{\mu}$:
\begin{align}
    &\hat{\mu}(M) = \frac{\bra{d-\hat{b}} P \ket{s_M}}{\bra{s_M} P \ket{s_M}}, \\
    &P = \mathds{1} - \Delta {(\Delta^{\intercal} \Delta)}^{-1} \Delta^{\intercal}. \nonumber
\end{align}

An interesting and important fact is that $P$ is a projector and it is symmetric:
\begin{align}
    &P^2 = P, \quad P = P^{\intercal}.
\end{align}

A projector is always positive semi-definite, which means that the product below is non-negative for any non-zero $\vec{s}$:
\begin{align}
    \bra{s} P \ket{s} = \bra{s} P^2 \ket{s} = {\left( P \ket{s} \right)}^2 \geq 0, \quad \forall \vec{s} \neq \vec{0} .\label{eq:denom}
\end{align}

Let us estimate the test statistic $t_M$:
\begin{align}
    &t_M = (-2 \ln \mathcal{L}_0) - (-2 \ln \mathcal{L}_1) \nonumber \\
    &\quad =  2 \braket{d - \hat{b} - \Delta \vec{\theta}_0}{\Delta \vec{\theta}_1 + \hat{\mu} s} \nonumber \\
    &\qquad + \braket{\Delta \vec{\theta}_1 + \hat{\mu} s}{\Delta \vec{\theta}_1 + \hat{\mu} s}.
\end{align}

We again use eq.~(\ref{eq:master-1}) to cancel the $\vec{\theta}_0$ contribution and eq.~(\ref{eq:nu-1}) to substitute the solution for $\vec{\theta}_1$:
\begin{align}
    &t_M = \hat{\mu} \bra{d-\hat{b}} P \ket{s_M} = \hat{\mu}^2 \bra{s_M} P \ket{s_M}. \label{eq:ts}
\end{align}
The significance $Z_M$, as defined in eq.~(\ref{eq:def-sig}), is:
\begin{align}
    &Z_M = \hat{\mu} \sqrt{\bra{s_M} P \ket{s_M}} = \frac{\bra{d-\hat{b}} P \ket{s_M}}{\sqrt{\bra{s_M} P \ket{s_M}}}. \label{eq:sig}
\end{align}
The square root in eq.~(\ref{eq:sig}) is always defined, as the product under the square root is always positive (eq.~(\ref{eq:denom})).

For the covariance matrix estimation, we would need to average over data. We are looking for a solution with uncorrelated fluctuations in each bin~(sec.~\ref{par:core-assumptions}), and we recall that we normalized the errors to $1$ in eq.~(\ref{eq:norm}), therefore, the following is true:
\begin{align}
    E_{d} \left\{ \ket{d-\hat{b}}\bra{d-\hat{b}} \right\} = \mathds{1}, \label{eq:d-mean}
\end{align}
where $E_d$ denotes the expectation value calculated across samples of the dataset.

The covariance matrix, then, is\footnotemark{}:
\begin{align}
    &\Sigma_{MN} = E_{d} \left\{ Z_M Z_N \right\} \nonumber \\
    &\quad = E_{d} \left\{ \frac{\bra{s_M} P \ket{d-\hat{b}}}{\sqrt{\bra{s_M} P \ket{s_M}}} \frac{\bra{d-\hat{b}} P \ket{s_N}}{\sqrt{\bra{s_N} P \ket{s_N}}} \right\} \nonumber \\
    &\quad = \frac{\bra{s_M} P }{\sqrt{\bra{s_M} P \ket{s_M}}} E_{d} \left\{\ket{d-\hat{b}}\bra{d-\hat{b}}\right\} \frac{ P \ket{s_N}}{\sqrt{\bra{s_N} P \ket{s_N}}} \nonumber \\
    &\quad = \frac{\bra{s_M}}{\sqrt{\bra{s_M} P \ket{s_M}}} P \frac{\ket{s_N}}{\sqrt{\bra{s_N} P \ket{s_N}}},
\end{align}
\footnotetext{%
To see the parallel with Demortier~\cite{CERN-2008-001},
one needs to think of the background model as a linear combination of vectors in $\Delta$.
Then eq.~(\ref{eq:master-2}) defines a vector $\ket{v_M} = \frac{P\ket{s_M}}{\sqrt{\bra{s_M}P\ket{s_M}}}$, which was introduced by Demortier and is orthogonal to each of the vectors constituting the background shape.
The test statistic, then, can be rewritten as $t_M = \left(\braket{d - \hat{b}}{v_M}\right)^2$,
and the covariance can be expressed as $\Sigma_{MN} = \braket{v_M}{v_N}$.
}
where we used the symmetry and projector properties of $P$.

It should be noted that from the data
fluctuations $\vec{d} - \vec{\hat{b}}$ contributing to the
covariance matrix in the form
\begin{align}
    Fluct. \propto E_{d} \left\{ \ket{d - \hat{b}}\bra{d - \hat{b}} \right\},
\end{align}
a superposition principle, relied on in ref.~\cite{ananiev2022}, can be derived:
\begin{align}
    \Sigma_{MN} = \sum_f \Sigma^{f}_{MN},
\end{align}
where $f$ enumerates independent fluctuations in different bins.

In summary, we can estimate the autocovariance matrix of the significance field from the signal model and derivatives of the background model:
\begin{align}\label{eq:final-cov}
    &\Sigma_{MN} = \frac{\bra{s_M}}{\sqrt{\bra{s_M} P \ket{s_M}}} P \frac{\ket{s_N}}{\sqrt{\bra{s_N} P \ket{s_N}}}, \quad M, N \in \mathcal{M} \nonumber \\
    &P = \mathds{1} - \Delta {(\Delta^{\intercal} \Delta)}^{-1} \Delta^{\intercal}, \\
    &\Delta_{i \alpha} = \frac{\partial b_i(\vec{\theta})}{\partial \theta^\alpha} \vert_{\vec{\theta} = \vec{0}}. \nonumber
\end{align}

\section{Justification of the set of Asimov background samples}\label{sec:asimov-set-just}

In this section we would like to compare the derived expression
eq.~(\ref{eq:final-cov}) for the linear approximation of the significance
covariance matrix
to the empirical study~\cite{ananiev2022} and the
$H \rightarrow \gamma\gamma$ inspired model introduced there.
To carry out the calculations we used the SigCorr package
that we developed specifically for trials factor studies,
which now includes functionality for the linear approximation~\cite{gitlab-sigcorr}.

\begin{figure}[ht]
    \centering
    \includegraphics[width=0.6\linewidth]{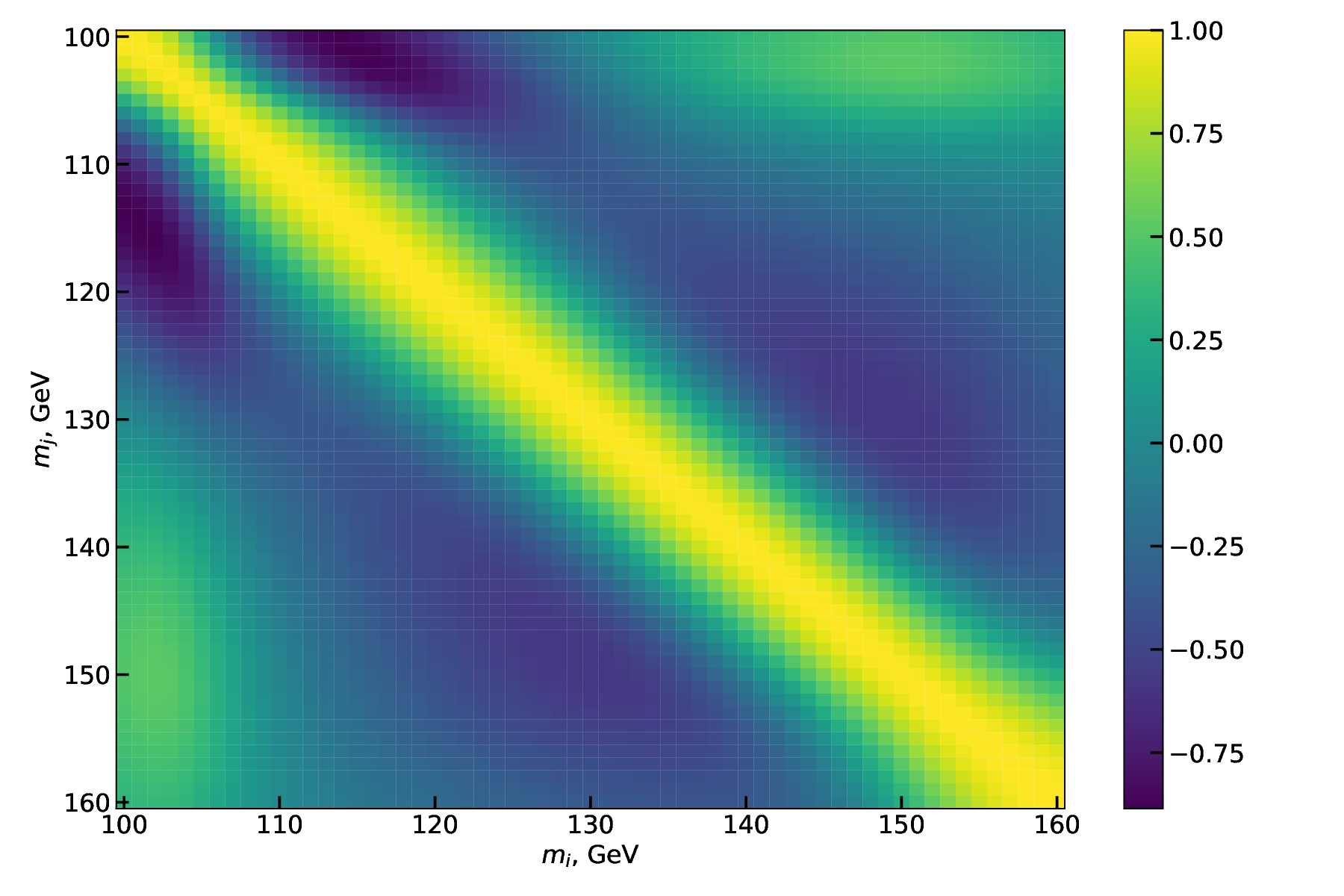}
    \caption{A linear approximation of the significance covariance matrix which was computed on the true parameters of the $H \rightarrow \gamma\gamma$ inspired model.}\label{fig:hyy-linasimov-cov-full}
\end{figure}

We estimate the linear approximation using eq.~(\ref{eq:final-cov})
with the true parameters of the model, which were predefined in the paper.
The resulting matrix shown in figure~\ref{fig:hyy-linasimov-cov-full}
is visually indistinguishable from the one presented in the empirical study.

\begin{figure}[ht]
    \centering
    \includegraphics[width=0.6\linewidth]{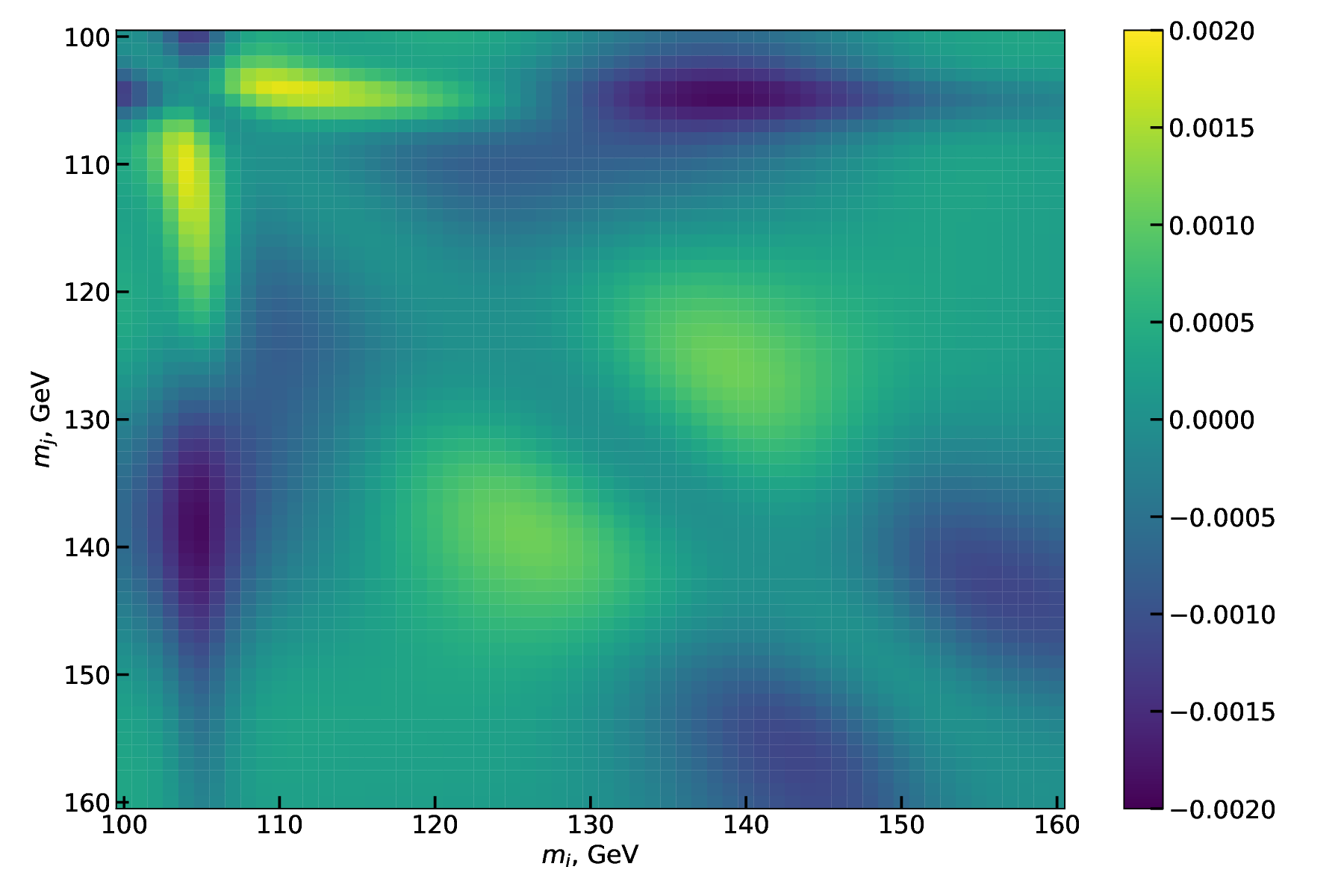}
    \caption{The difference between the linear approximation of the significance covariance matrix computed with the true parameters of the $H \rightarrow \gamma\gamma$ inspired model (figure~\ref{fig:hyy-linasimov-cov-full}) and the covariance matrix estimated with the set of Asimov background samples~\cite{ananiev2022}.}\label{fig:hyy-linasimov-covcmp_asimov-diff}
\end{figure}

We also show, in figure~\ref{fig:hyy-linasimov-covcmp_asimov-diff},
the difference between the linear approximation computed on
the model's true parameters (figure~\ref{fig:hyy-linasimov-cov-full})
and the empirical estimate.
We confirm that the empirical covariance matrix is compatible with
the linear approximation suggested in this paper
within the accuracy of the empirical estimate.

On the one hand, the compatibility of the linear approximation and
the empirical study allows us to refer to the validations conducted in
the empirical study, including those regarding trials factor estimation,
and to re-apply them to the method suggested in this paper.
The direct calculation of the up-crossings from the covariance matrix, described in~\cite{ananiev2022}, becomes particularly appealing now, since it requires only a single fit of the statistical model to the data.

The linear approximation, on the other hand, serves as the theoretical basis
for the empirical set of Asimov background samples used to estimate the covariance matrix in the aforementioned work.

\section{Conclusion}
In this work we proposed a novel method for the estimation of the covariance matrix of statistical
significance in new particle searches using a linear expansion of the statistical
model around its background-only best fit to the data.
In addition to the closed form expression for the linear approximation
of the significance covariance matrix,
we also presented elegant expressions for the best fitted signal strength
and statistical significance in this approximation.

We proved that the suggested covariance matrix satisfies the superposition
principle with regard to the fluctuations of the data, which makes it a good
proxy to the covariance matrix constructed with
the set of Asimov background samples\cite{ananiev2022}.

Finally, we compared these two approaches with
the example of a $H \rightarrow \gamma\gamma$ inspired model
and showed that the deviations are compatible with the error of
the set of Asimov background samples.

We, therefore, claim that all the validations conducted in
the empirical study, including those regarding trials factor estimation,
hold for the linear approximation suggested in this paper,
and the linear approximation serves as a theoretical basis for
the empirical set of Asimov background samples construction.

\acknowledgments{}

We would like to thank Elliot Reynolds for the encouraging discussion at the HDBS Workshop at Uppsala.
This research was supported by
the European Union Framework Programme for Research and Innovation Horizon 2020 (2014--2021)
under the Marie Sklodowska-Curie Grant Agreement No.765710.

\bibliographystyle{JHEP}
\bibliography{bibliography}   

\end{document}